\documentclass[twocolumn,showpacs,preprintnumbers,amsmath,amssymb]{revtex4}
\usepackage{graphicx}
\usepackage{bm}
\usepackage[dvipsnames]{color}
\begin{document}
\def \brho{{\hbox{\boldmath $\rho$}}}
\def \beps{{\hbox{\boldmath $\epsilon$}}}
\def \bdelta{{\hbox{\boldmath $\delta$}}}
\title{Changing character of electronic transitions in graphene: From single particle excitations to plasmons}
\author{Dino Novko$^{1}$}
\author{Vito Despoja$^{1,2,3}$}
\author{Marijan \v Sunji\' c$^{1,2}$}
\affiliation{$^1$Donostia International Physics Center (DIPC), P. Manuel de Lardizabal, 20018 San Sebastian, Basque Country, Spain}
\affiliation{$^2$Department of Physics, University of Zagreb, Bijeni\v{c}ka 32, HR-10000 Zagreb, Croatia}
\affiliation{$^3$Universidad del Pais Vasco, Centro de Fisica de Materiales CSIC-UPV/EHU-MPC, Av. Tolosa 72, E-20018 San Sebastian, Spain}
\begin{abstract}
In this paper we clarify the nature of $\pi$ and $\pi+\sigma$ electron excitations in pristine 
graphene. We clearly demonstrate the continuous transition from single particle to collective 
character of such excitations and how screening modifies their dispersion relations. We prove that $\pi$ and $\pi+\sigma$ plasmons do exist in graphene, 
though occurring only for a particular range of wavevectors and with finite damping rate. The particular attention is paid to compare the theoretical results with available EELS measurements in optical ($\mathrm{Q\approx 0}$) and other ($\mathrm{Q\neq 0}$) limits. The conclusions, based on microscopic numerical results, are confirmed in an approximate analytical approach.
\end{abstract}

\maketitle

\section{\label{sec:1} Introduction}
Plasmon spectra of a pristine single layer graphene were first obtained in Ref. \cite{firstpi}, where the authors observed two distinct structures which they attributed to the so-called $\pi$ and $\pi+\sigma$ plasmons. They observed that these two plasmonic modes were redshifted in comparison to the corresponding modes in the bulk graphite \cite{graphite00,graphite0,graphite1}, due to the reduction of macroscopic screening when going from graphite to graphene \cite{graphite2}. The early momentum-dependent theoretical and experimental measurements observed linear dispersion of this $\pi$ plasmon in graphene \cite{lint1,lint2,line1,line2}, which differs from the $\mathrm{Q^2}$ dispersion reported in graphite \cite{graphite2,graphite00,graphite0,graphite3}.

Recently a resolute claim was made \cite{Nanolett} that the previously accepted attribution of the two strong structures in 
the graphene excitation spectra was wrong, and the $\pi$ and $\pi+\sigma$ plasmons are in fact strong single particle (SP) $\pi\rightarrow\pi^*$ and $\sigma\rightarrow\sigma^*$ excitations, respectively, with a characteristic $\mathrm{Q^2}$ excitation energy dependence. Another group \cite{Pipl1} found strong evidence for 2D plasmon character of $\pi$ and $\sigma$ electron excitations, based on the electron energy loss spectroscopy (EELS) experiment showing the $\sqrt{\mathrm{Q}}$ dependent dispersion. Even taking into account possible uncertainties arising from experimental difficulties in EELS measurements for low Q values, it is obvious that this apparent controversy deserves to be analysed and resolved.

In this paper we solve this problem using both a numerical method and analytic arguments, providing a rigorous method of determining collective vs. single particle excitation character in solids, and apply it to analyse $\pi$ and $\sigma$ electron excitations in a self-supporting monolayer of pristine graphene. We find that the character of $\pi\rightarrow\pi^{\ast}$ (and $\sigma\rightarrow\sigma^{\ast}$) transition changes, depending on the wavevector Q. For small $\mathrm{Q}\approx 0$ these are unscreened single particle transitions, but with increasing Q they acquire collective character as the dynamical screening mechanism becomes more efficient. This explains the gradual change from the Q$^2$ dependence of excitation energies near $\mathrm{Q}\approx 0$ to the quasi-linear dependence at larger Q. And finally, for even larger Q the collective nature of this modes in graphene is suppressed and they again emerge as the single particle excitations. Same kind of dispersion is observed in \cite{Pipl21,Pipl2,duncan}, but the authors did not analyse it in detail. Although the described dispersion seems like the characteristic $\sqrt{\mathrm{Q}}$ dependence of the 2D plasmon, we show that this cannot be true because of the complex nature of this mode.

The described analysis is quite general, and its application to graphene provides a very nice illustration how an electronic process can change its character from an interband single particle transition to a collective mode as the dynamical screening takes over with the increasing wavevector $\mathrm{Q}$. 

In Sec. \ref{sec:2} we describe the derivation of the electronic excitation spectra $\mathrm{S(\mathrm{Q},\omega)}$ in terms of the dielectric tensor ${\cal E}_{\mathbf{GG'}}(\mathbf{Q},\omega)$, using the method of Ref. \cite{Gr2013}, and define the dynamical screening factor $\mathrm{D(\mathbf{Q},\omega)}$. In Sec. \ref{sec:3} we calculate numerically the macroscopic dielectric functions $\mathrm{Im\ {\cal E}_M(\mathbf{Q},\omega)}$, $\mathrm{Re\ {\cal E}_M(\mathbf{Q},\omega)}$ and $-\mathrm{Im\ 1/{\cal E}_M(\mathbf{Q},\omega)}$, as well as the dynamical screening factor $\mathrm{D(\mathbf{Q},\omega)}$, and discuss the excitation spectra. In Sec. \ref{sec:4} we summarize the results and their relation to previous experimental and theoretical work.

\section{\label{sec:2} Formulation of the problem}
The first part of the calculation consists of determining the Kohn-Sham (KS) ground state of graphene and the corresponding wave functions and energies. For the unit cell constant we use the experimental value of $a=4.651\ \mathrm{a.u.}$ \cite{lattice}, and we separate the graphene layers with the distance $L=5a$. For calculating KS wave functions and energies we use a plane-wave self-consistent field DFT code (PWSCF) within the QUANTUM ESPRESSO (QE) package \cite{QE}. The core-electron interaction was approximated by the norm-conserving pseudopotentials \cite{normcon}, and the exchange correlation (XC) potential by the Perdew-Zunger local density approximation (LDA) \cite{lda1}. To calculate the ground state electronic density we use $30\times30\times1$ Monkhorst-Pack K-point mesh of the first Brillouin zone (BZ) and for the plane-wave cut-off energy we choose 50 Ry.

Using the wave functions and energies obtained in the described way we perform the calculation of the electronic excitation spectra within the random phase approximation (RPA). In the quasi-2D systems such as graphene, with the electronic density in the region $0<z<L$, the spectral function of electronic excitations can be defined as \cite{Gr2013}
\begin{equation}
\mathrm{S(\mathbf{Q},\omega)}=-\mathrm{Im}\left[{\cal E}_{\mathbf{GG'}}^{-1}(\mathbf{Q},\omega)\right]_{\mathbf{G=G'=0}},
\label{spec1}
\end{equation}
where the dielectric matrix in the RPA is given by 
\begin{equation}
{\cal E}_{\textbf{G}\textbf{G}'}(\textbf{Q},\omega)=
\delta_{\textbf{G}\textbf{G}'}-
\sum_{\textbf{G}_{1}}V_{\textbf{G}\textbf{G}_{1}}(\textbf{Q})\chi^{0}_{\textbf{G}_{1}\textbf{G}'}(\textbf{Q},\omega).
\label{epsm}
\end{equation}
The noninteracting charge-charge response function is given in matrix form as
\begin{eqnarray}
&\chi^{0}_{{\bf G}{\bf G}'}({\bf Q},\omega)=\frac{2}{V}\sum\limits_{{\bf K},n,m}\ \frac{f_n({\bf K})-f_m({\bf K}+{\bf Q})}{\omega+i\eta+E_n({\bf K})-E_m({\bf K}+{\bf Q})} \nonumber \\
&\times M_{n{\bf K},m{\bf K}+{\bf Q}}({\bf G})\ M^*_{n{\bf K},m{\bf K}+{\bf Q}}({\bf G'}).
\label{Resfun0}
\end{eqnarray}
The $V=S\times L$ is the normalization volume, $S$ is the normalization surface and $f_n({\bf K})=\theta[E_F-E_n({\bf K})]$ is the Fermi-Dirac distribution at $T=0$. In this summation we use $201\times201\times1$ K-mesh sampling and up to 70 electronic bands. For the broadening parameter $\eta$ we use 0.05 eV.
Matrix elements in (\ref{Resfun0}) have the form  
\begin{equation}
M_{n{\bf K},m{\bf K}+{\bf Q}}({\bf G})=\left\langle \Phi_{n{\bf K}}\left|e^{-i({\bf Q}+{\bf G}){\bf r}}\right|\Phi_{m{\bf K}+{\bf Q}}\right\rangle_V,
\label{Matrel}
\end{equation}
where ${\bf Q}$ is the momentum transfer vector parallel to the $x$-$y$ plane, ${\bf G}=({\bf G}_\parallel,G_z)$ are 3D reciprocal lattice vectors and ${\bf r}=(\brho,z)$ is a 3D position vector. Wave functions $\Phi_{n{\bf K}}(\mathbf{r})$ are KS wave functions from the ground state calculation and $E_n({\bf K})$ are the corresponding energies.
\begin{figure}[b]
\includegraphics[width=\columnwidth]{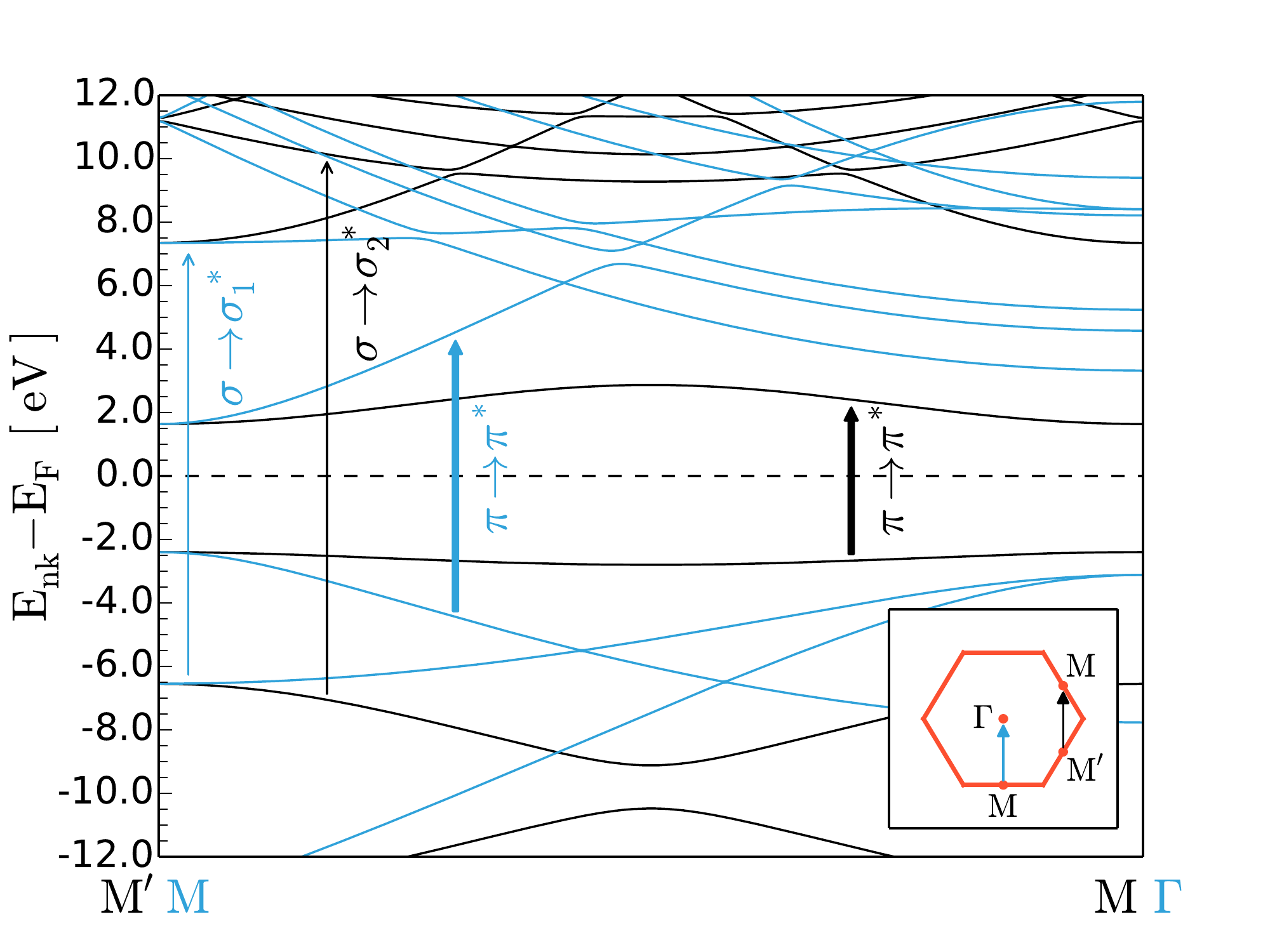}
\caption{Main panel: Graphene band structure along the $\mathrm{M'-M}$ (black lines) and $\mathrm{M-\Gamma}$ (blue lines) directions of the BZ. Interband transitions between $\pi$ bands are marked with thick arrows, while the $\sigma$ band transitions are marked with thin arrows. Inset: BZ of graphene with black and blue arrows showing directions of corresponding band structures in the main panel.}
\label{Fig2}
\end{figure}
In this approach the superlattice consists of periodically repeated layers of graphene such that the charge densities of adjacent layers do not overlap. We restrict our consideration to one layer placed in the region $0<z<L$, where the interaction with the adjacent layers is avoided by allowing Coulomb interaction between charge oscillations only within this region. This is done by integrating the Dyson equation for $\chi$ within the limits of $0<z<L$ \cite{Gr2013,Osc1,Osc2}. The resulting Coulomb interaction matrix elements have the explicit form 
\begin{eqnarray}
&V_{\textbf{G}_{1}\textbf{G}_{2}}(\textbf{Q})=\frac{4\pi}{\left|\textbf{Q}+\textbf{G}_1\right|^2}\delta_{\textbf{G}_1\textbf{G}_2}-p_{G_{z1}}p_{G_{z2}}
\frac{4\pi(1-e^{-\left|\textbf{Q}+\textbf{G}_{\parallel1}\right|L})}
{\left|\textbf{Q}+\textbf{G}_{\parallel1}\right|L}\nonumber \\
&\times\frac{\left|\textbf{Q}+\textbf{G}_{\parallel1}\right|^2-{G}_{z1}{G}_{z2}}
{(\left|\textbf{Q}+\textbf{G}_{\parallel1}\right|^2+{G}^2_{z1})
(\left|\textbf{Q}+\textbf{G}_{\parallel1}\right|^2+{G}^2_{z2})}
\delta_{\textbf{G}_{\parallel1}\textbf{G}_{\parallel2}}
\label{3Dexm}
\end{eqnarray}
with
\[
p_{G_z}=
\left\{\begin{array}{ccc}
1;&\ {G_z}=\frac{2k\pi}{L}&
\\
-1;&\ {G_z}=\frac{(2k+1)\pi}{L}&,\ \ k=0,1,2,3,..
\end{array}
\right.
\]
Similar approach was also carried out for layered structures in Ref. \cite{similar}.

In our analysis of electronic excitation spectra we will use the macroscopic dielectric function which is defined as

\begin{equation}
\mathrm{{\cal E}_M(\mathbf{Q},\omega)}=\frac{1}{\left[{\cal E}_{\mathbf{GG'}}^{-1}(\mathbf{Q},\omega)\right]_{\mathbf{G=G'=0}}},
\end{equation}
and includes the crystal local field effects in the perpendicular, though not in the parallel direction. By this we mean that we put $\mathrm{G}_{||}=0$, while leaving the reciprocal lattice vectors in $z$ direction, $G_z$, which is justified by the fact that  the parallel local field effects are not so important for describing the surface plasmons \cite{LFE1}. To get a well converged spectra we use 71 $G_z$ vectors.

With this macroscopic dielectric function the spectral function (\ref{spec1}) can be written as

\begin{eqnarray}
\mathrm{S(\mathbf{Q},\omega)}&=&\mathrm{\frac{{\cal E}_2(\mathbf{Q},\omega)}{{\cal E}_1^2(\mathbf{Q},\omega)+{\cal E}_2^2(\mathbf{Q},\omega)}}\nonumber \\
&=&\mathrm{{\cal E}_2(\mathbf{Q},\omega)D(\mathbf{Q},\omega)},
\label{spec2}
\end{eqnarray}
where we have defined the dynamical screening factor $\mathrm{D(\mathbf{Q},\omega)}$ and simplified the notation with $\mathrm{{\cal E}_1\equiv Re\ {\cal E}_M}$ and $\mathrm{{\cal E}_2\equiv Im\ {\cal E}_M}$. For a vanishing screening in the system we have that $\mathrm{D}\rightarrow 1$, while for the screened $\mathrm{D}\neq1$. In the first case the spectral function S is equal to $\mathrm{{\cal E}_2}$ and all the structures in spectra have purely SP character.

\begin{figure*}[t]
\includegraphics[width=0.32\textwidth]{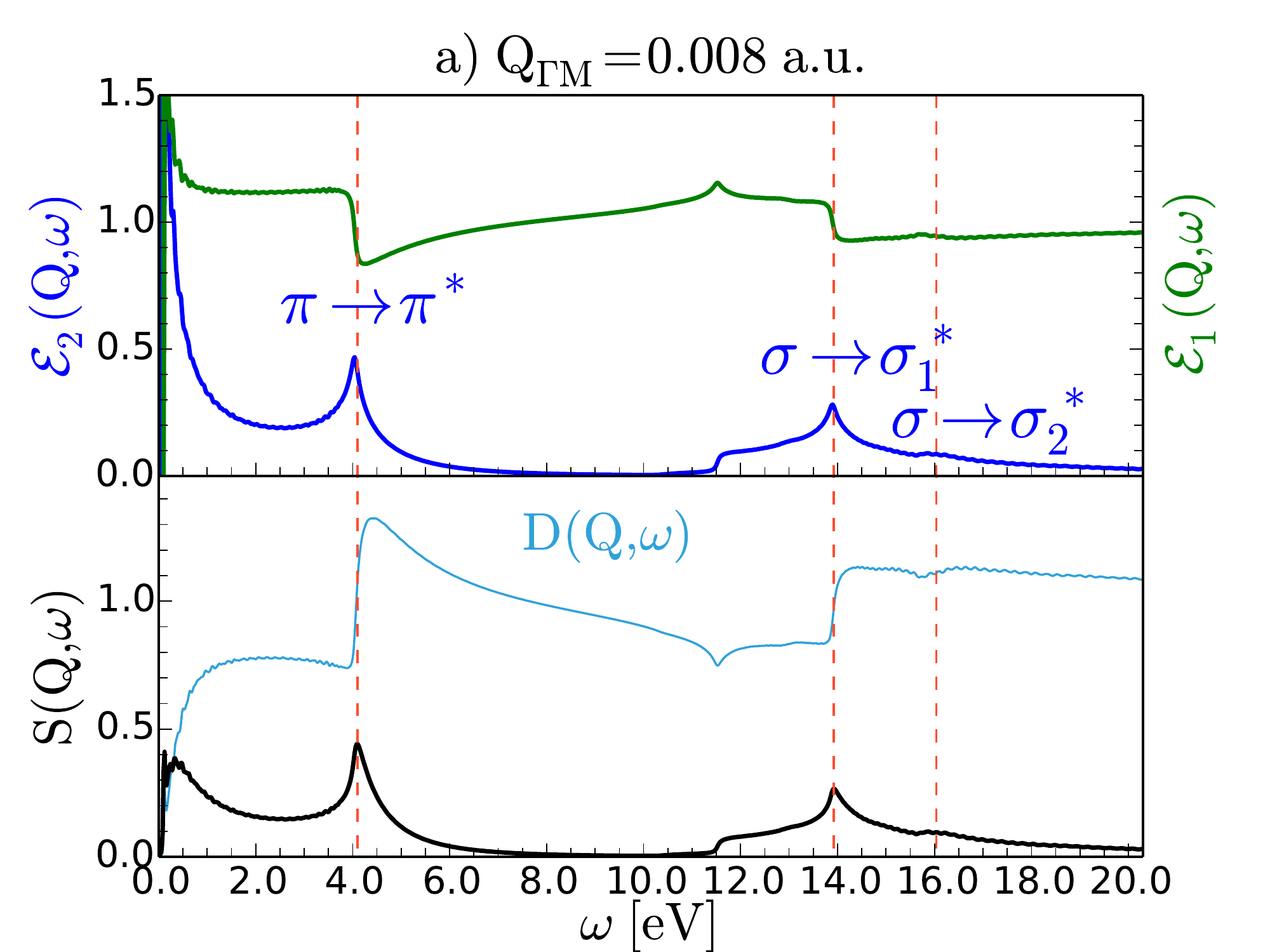}
\includegraphics[width=0.32\textwidth]{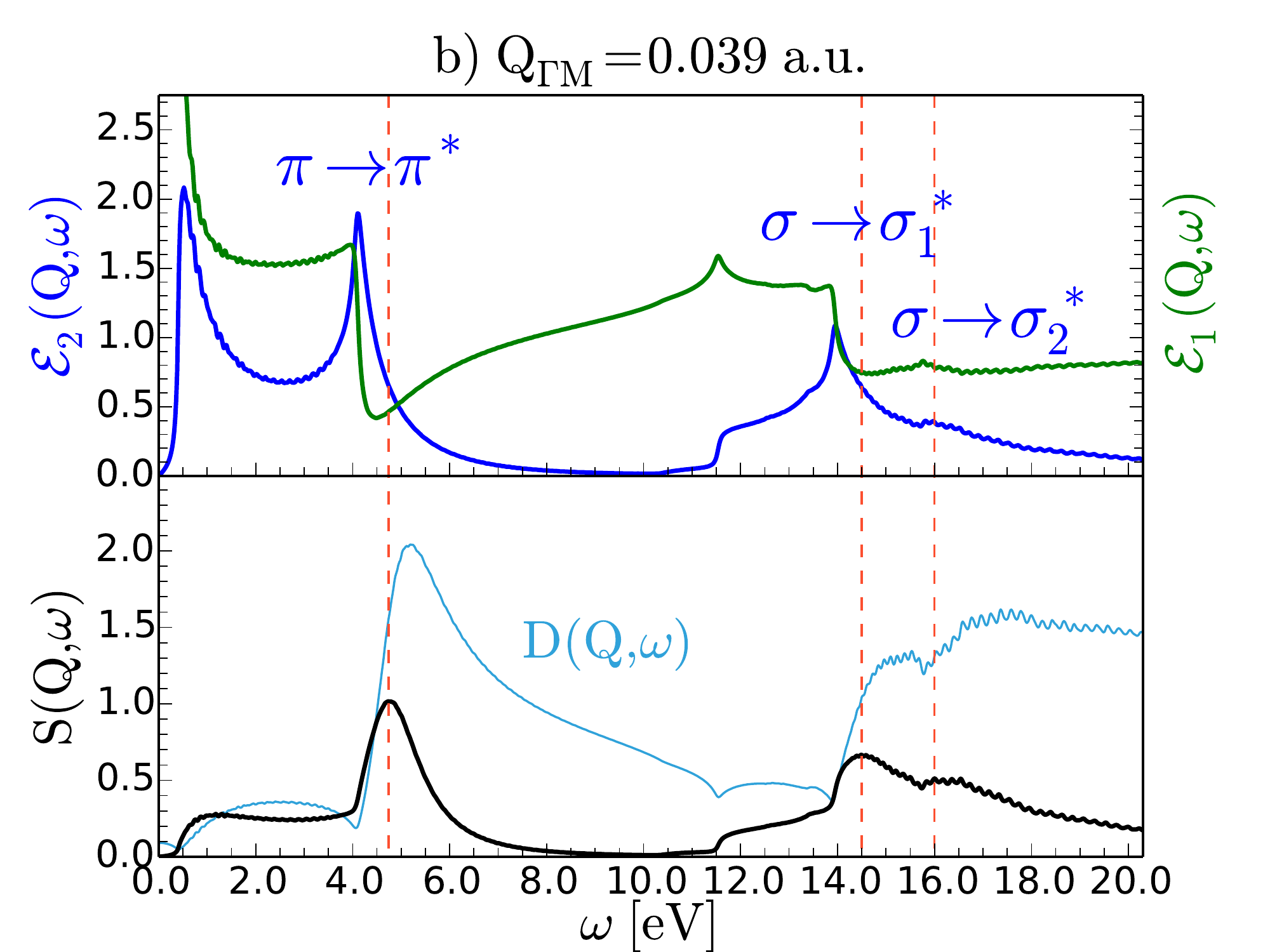}
\includegraphics[width=0.32\textwidth]{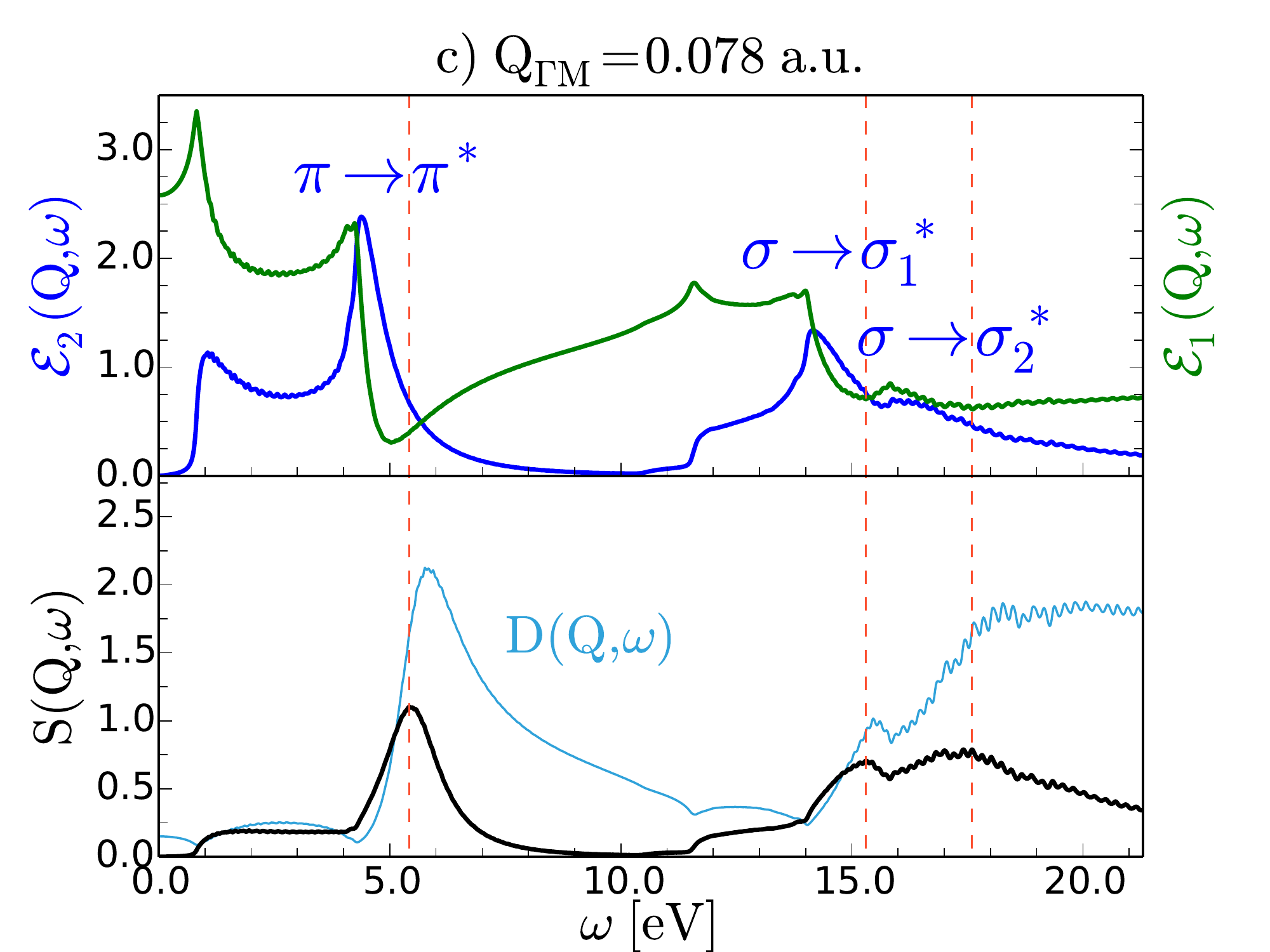}
\includegraphics[width=0.32\textwidth]{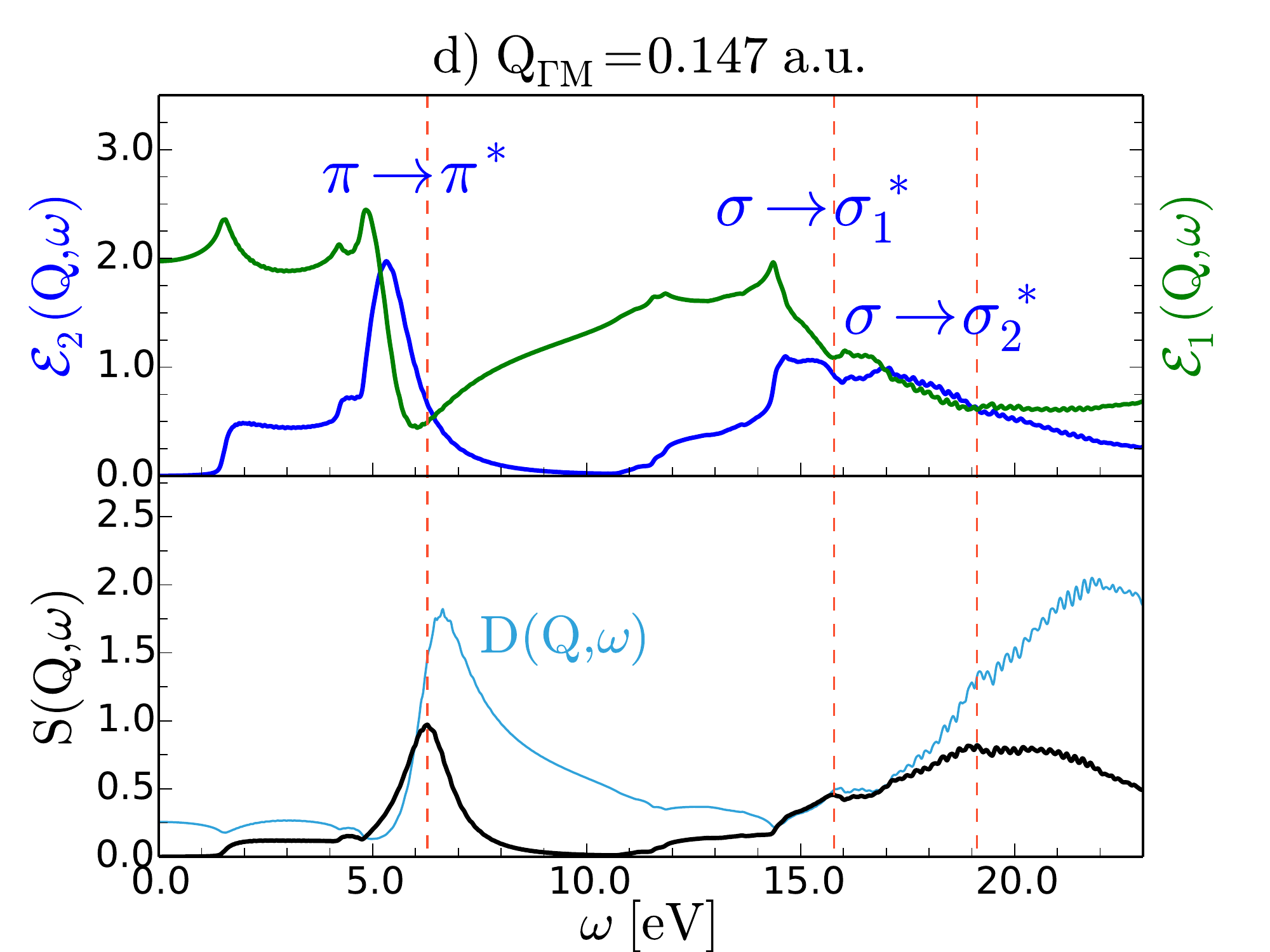}
\includegraphics[width=0.32\textwidth]{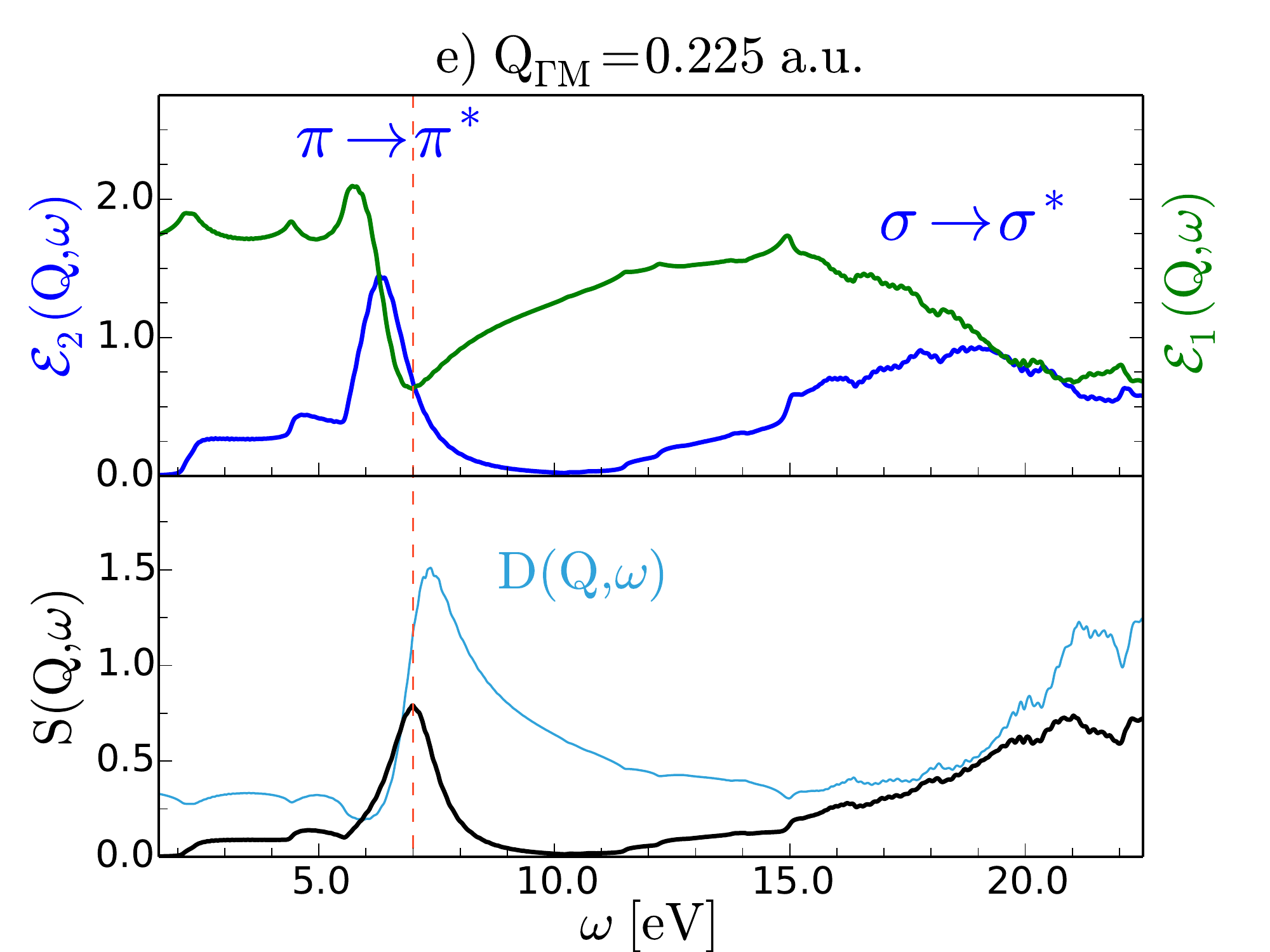}
\includegraphics[width=0.32\textwidth]{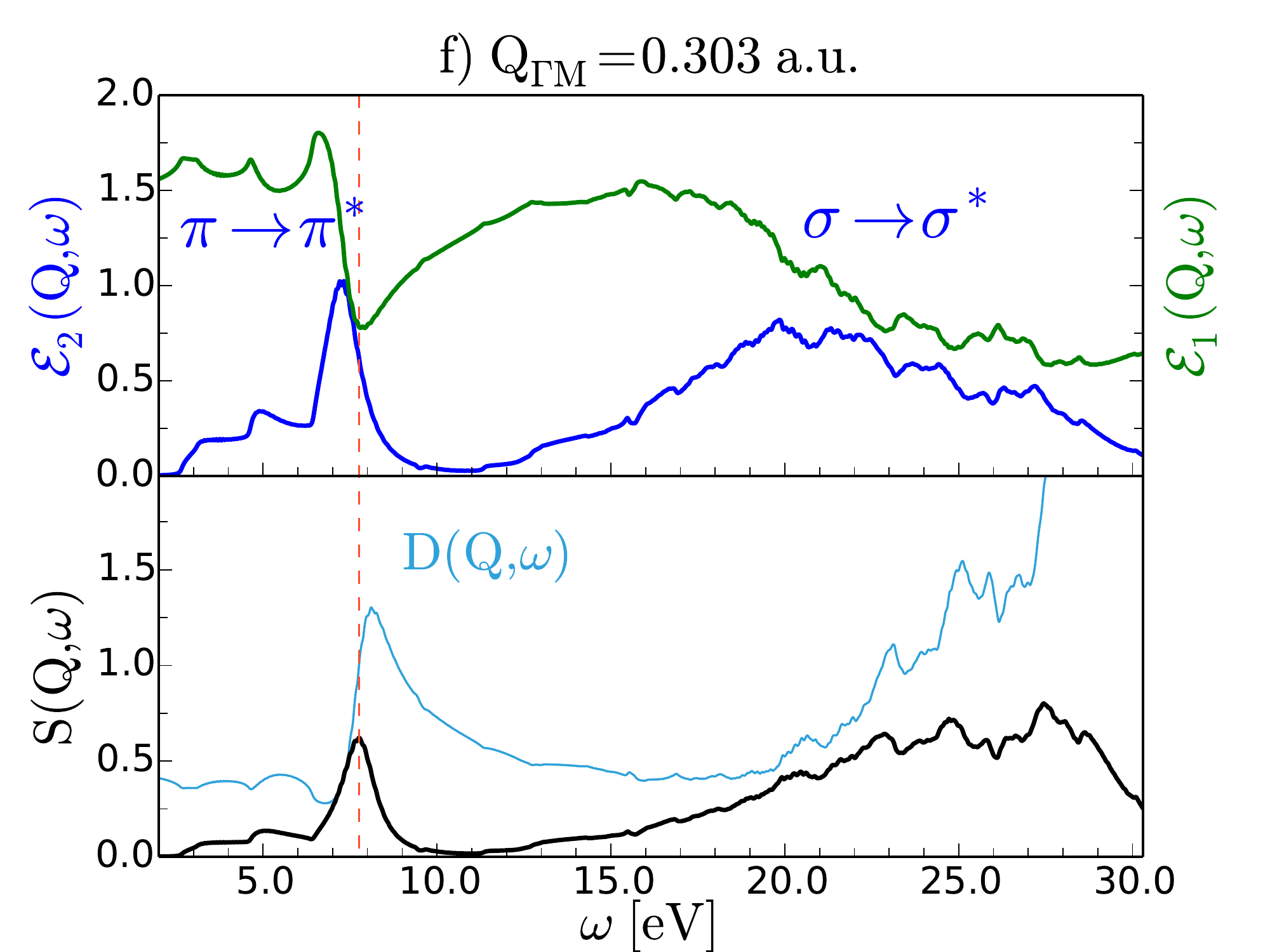}
\caption{Real (green) and imaginary (blue) parts of the macroscopic dielectric function $\mathrm{{\cal E}_M(Q,\omega)}$, dynamical screening factor $\mathrm{D(Q,\omega)}$ (light blue) and spectra of electronic excitations (black) in pristine graphene for six different Q vectors along the $\mathrm{\Gamma M}$ direction: a) $\mathrm{Q_{\Gamma M}=0.008\ a.u.}$, b) $\mathrm{Q_{\Gamma M}=0.039\ a.u.}$, c) $\mathrm{Q_{\Gamma M}=0.078\ a.u.}$, d) $\mathrm{Q_{\Gamma M}=0.147\ a.u.}$, e) $\mathrm{Q_{\Gamma M}=0.225\ a.u.}$, f) $\mathrm{Q_{\Gamma M}=0.303\ a.u.}$. Red vertical dashed lines denote the energy positions of $\pi$, $\sigma_1$ and $\sigma_2$ plasmons for each Q vector.} 
\label{Fig1}
\end{figure*}

To analyse the electronic excitation spectrum of a 2D material one can also use the dielectric function within the tight-binding approximation (TBA) \cite{Kupcic,tba1,tba2}. Instead of the KS wave functions and energies, here one uses the states and energies of the TBA hamiltonian. If we consider graphene beyond the Dirac cone approximation then the 2D charge-charge response function is given by

\begin{eqnarray}
&\chi^{0}_{\mathrm{TBA}}({\bf Q},\omega)=\frac{2}{S}\sum\limits_{{\bf K},\mu,\mu'}\ \frac{f_{\mu}({\bf K})-f_{\mu'}({\bf K}+{\bf Q})}{\omega+i\eta+E_{\mu}({\bf K})-E_{\mu'}({\bf K}+{\bf Q})} \nonumber \\
&\times \left| \frac{1}{2}\left( 1+\mu\mu'\frac{g^{\ast}(\mathbf{K})g(\mathbf{K+Q})}{|g(\mathbf{K})||g(\mathbf{K+Q})|} \right) \right|^2,
\label{TBchi}
\end{eqnarray}
where the band index $\mu=-1$ represents the occupied $\pi$ band and $\mu=1$ the unoccupied $\pi^{\ast}$ band in graphene. For the numerical calculation of (\ref{TBchi}) we use $600\times600\times1$ K-point mesh and the broadening parameter $\eta=0.05\ \mathrm{eV}$. The TBA band energies are $E_{\mu}({\bf K})=\mu\gamma|g(\mathbf{K})|$ with the hopping parameter $\gamma\approx2.02\ \mathrm{eV}$ \cite{gamma0,gamma1}, while the hopping function is given by

$$
g(\mathbf{K})=e^{i\mathrm{K_y}a/\sqrt{3}}+2e^{i\mathrm{K_y}a/2\sqrt{3}}\cos(\mathrm{K_x}a/2),
$$
where $a$ is the lattice constant of graphene. Here the spectral function is also defined as in (\ref{spec1}) and 
(\ref{spec2}), but the dielectric function does not have crystal local field effects included and can be written as 
\begin{equation}
\varepsilon({\bf Q},\omega)=1-\frac{2\pi}{\mathrm{Q}}\chi^0_{\mathrm{TBA}}({\bf Q},\omega).
\label{TBeps}
\end{equation}

\section{\label{sec:3} Results and discussion}
In order to better understand the spectrum of SP electronic excitations we will first analyze the graphene band structure. Fig. \ref{Fig2} shows the graphene band structure along the $\mathrm{M'-M}$ (black lines) and $\mathrm{M-\Gamma}$ (blue lines) directions of the BZ, which are relevant directions when the wavevector of external perturbation, Q, is in the $\mathrm{\Gamma M}$ direction. We see that $\pi$ electrons exhibit two different kinds of interband transitions. The first is attributed to transitions between two almost dispersionless $\pi$ bands along the $\mathrm{M'-M}$ direction, as denoted by black thick arrow in Fig. \ref{Fig2}. The second kind of $\pi$ interband transitions are attributed to transitions along the $\mathrm{M-\Gamma}$ direction, as denoted by a 
thick blue arrow in Fig. \ref{Fig2}. We shall see that the $\pi$ plasmon can be formed from the latter transitions when they are dynamically screened. Two other transitions are between occupied and unoccupied $\sigma$ bands. They can be divided into $\sigma\rightarrow\sigma_1^{\ast}$ and $\sigma\rightarrow\sigma_2^{\ast}$  transitions \cite{sigma1}, as denoted by thin blue and black arrows in Fig. \ref{Fig2}. Moreover, the $\sigma_1$ and $\sigma_2$ plasmons, usually treated as one $\pi+\sigma$ plasmon, originate from these transitions. 

\begin{figure*}[t]
\includegraphics[width=0.32\textwidth]{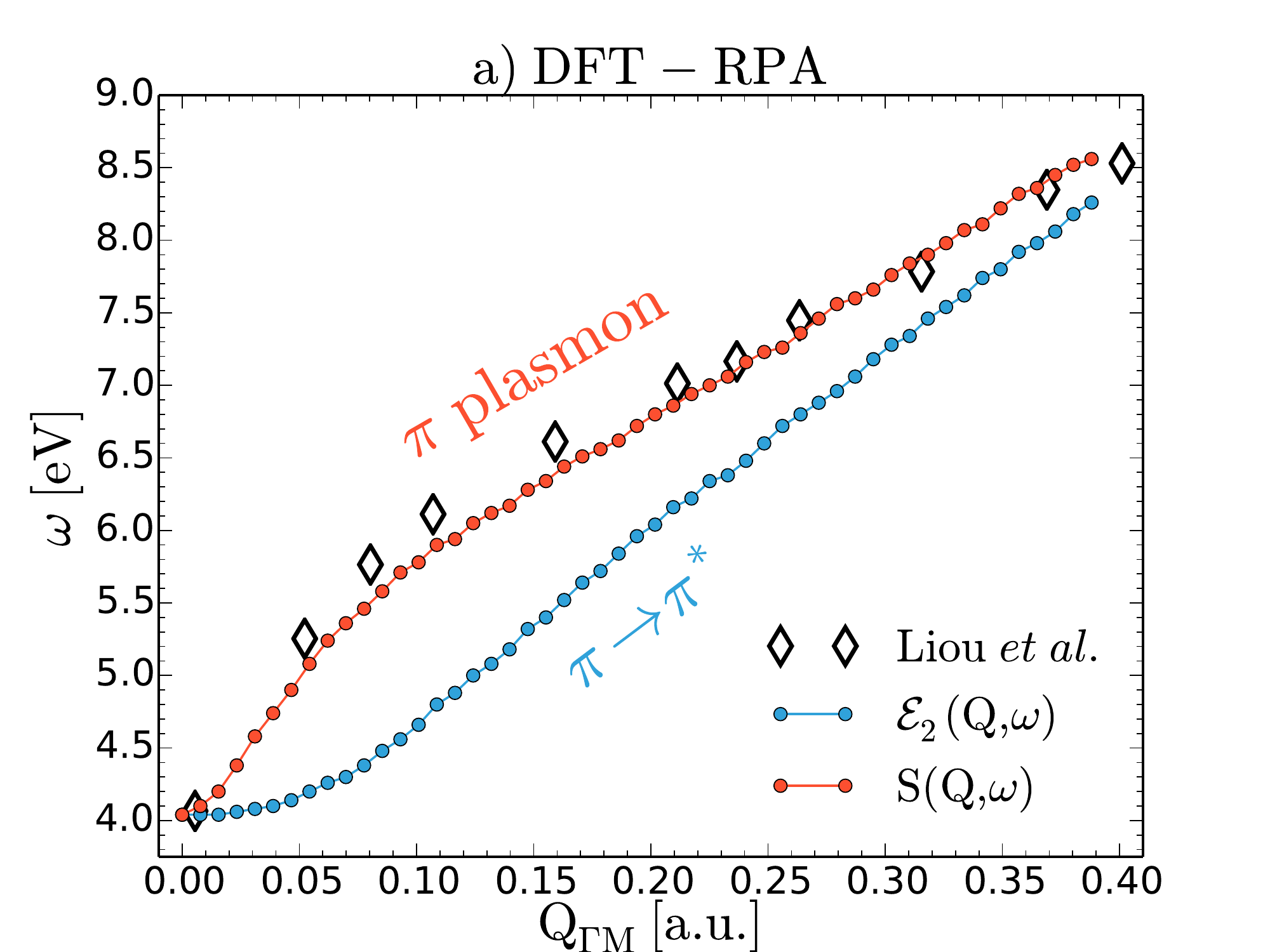}
\includegraphics[width=0.32\textwidth]{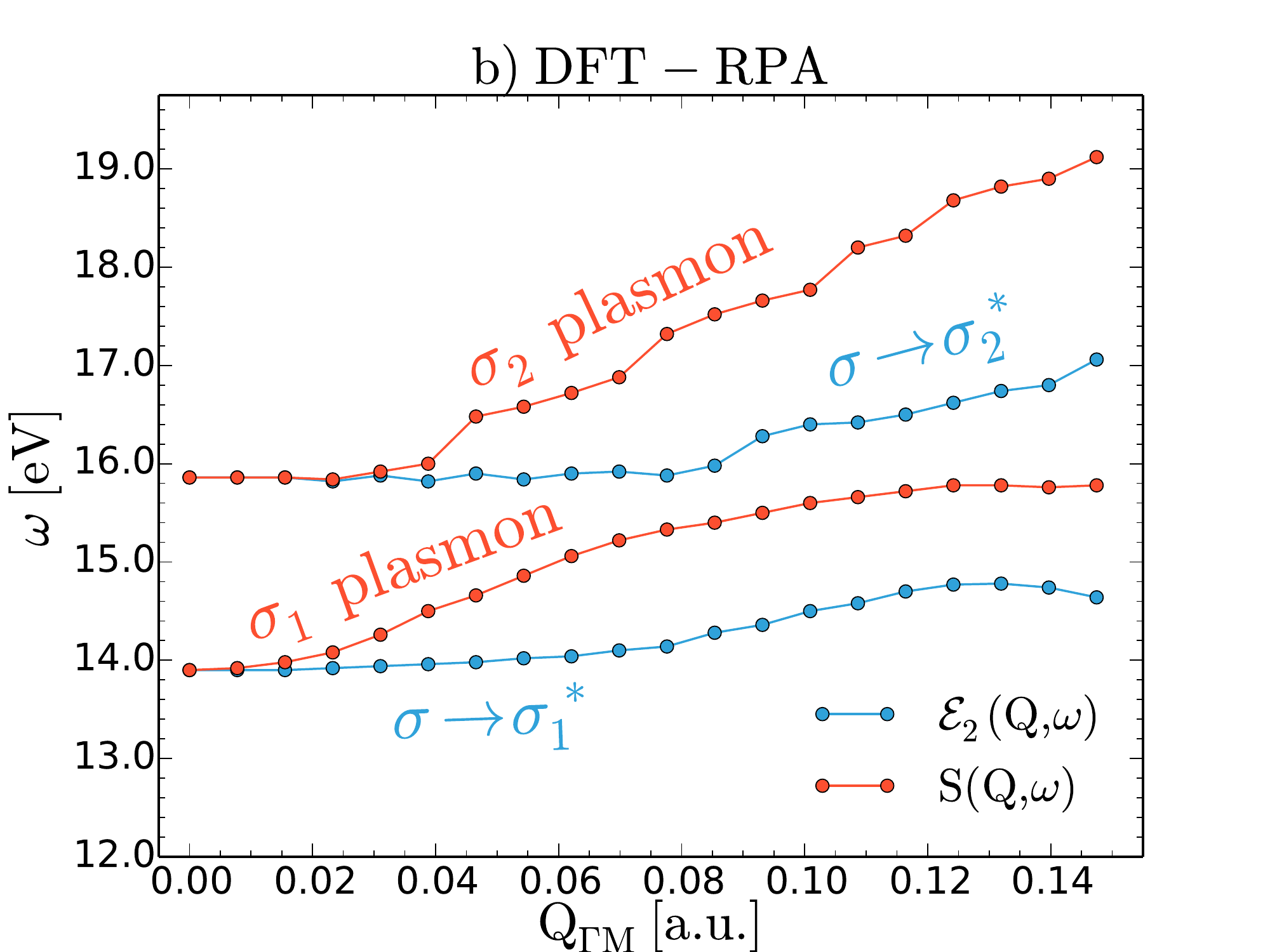}
\includegraphics[width=0.32\textwidth]{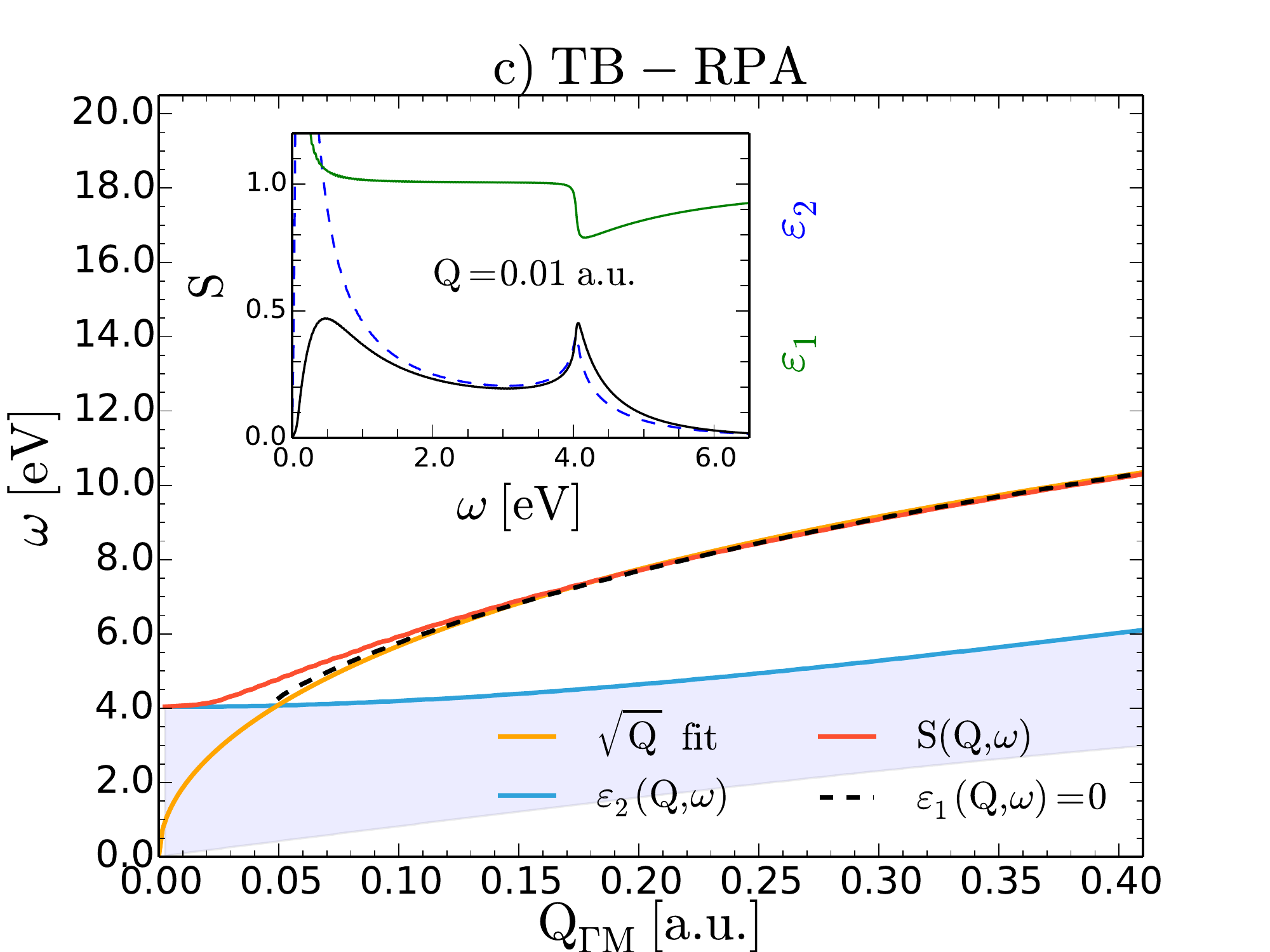}
\caption{a) Energies of $\pi\rightarrow\pi^{\ast}$ interband transitions (blue) and the $\pi$ plasmon (red) as functions of the wavevector Q along $\mathrm{\Gamma M}$ direction obtained from DFT-RPA method. Black unfilled diamonds represent the experimental data from Ref. \cite{Pipl1}. b) Same as in a) for $\sigma\rightarrow\sigma_1^{\ast}$ and $\sigma\rightarrow\sigma_2^{\ast}$ (blue) transitions and $\sigma_1$ and $\sigma_2$ plasmons (red). The point $\mathrm{Q=0}$ is treated separately \cite{q0}. c) Same as in a) but showing the results obtained with the TB-RPA method. In addition the points where $\mathrm{\varepsilon_1(Q,\omega)=0}$ are shown by the black dashed line and the corresponding $\sqrt{\mathrm{Q}}$ fit with the orange line. Inset: Real (green) and imaginary (dashed blue) parts of $\mathrm{\varepsilon(Q,\omega)}$ and $\mathrm{S(Q,\omega)}$ (black) for $\mathrm{Q=0.01\ a.u.}$ as obtained with TB-RPA method.}
\label{Fig3}
\end{figure*}

\begin{figure}[b]
\includegraphics[width=\columnwidth]{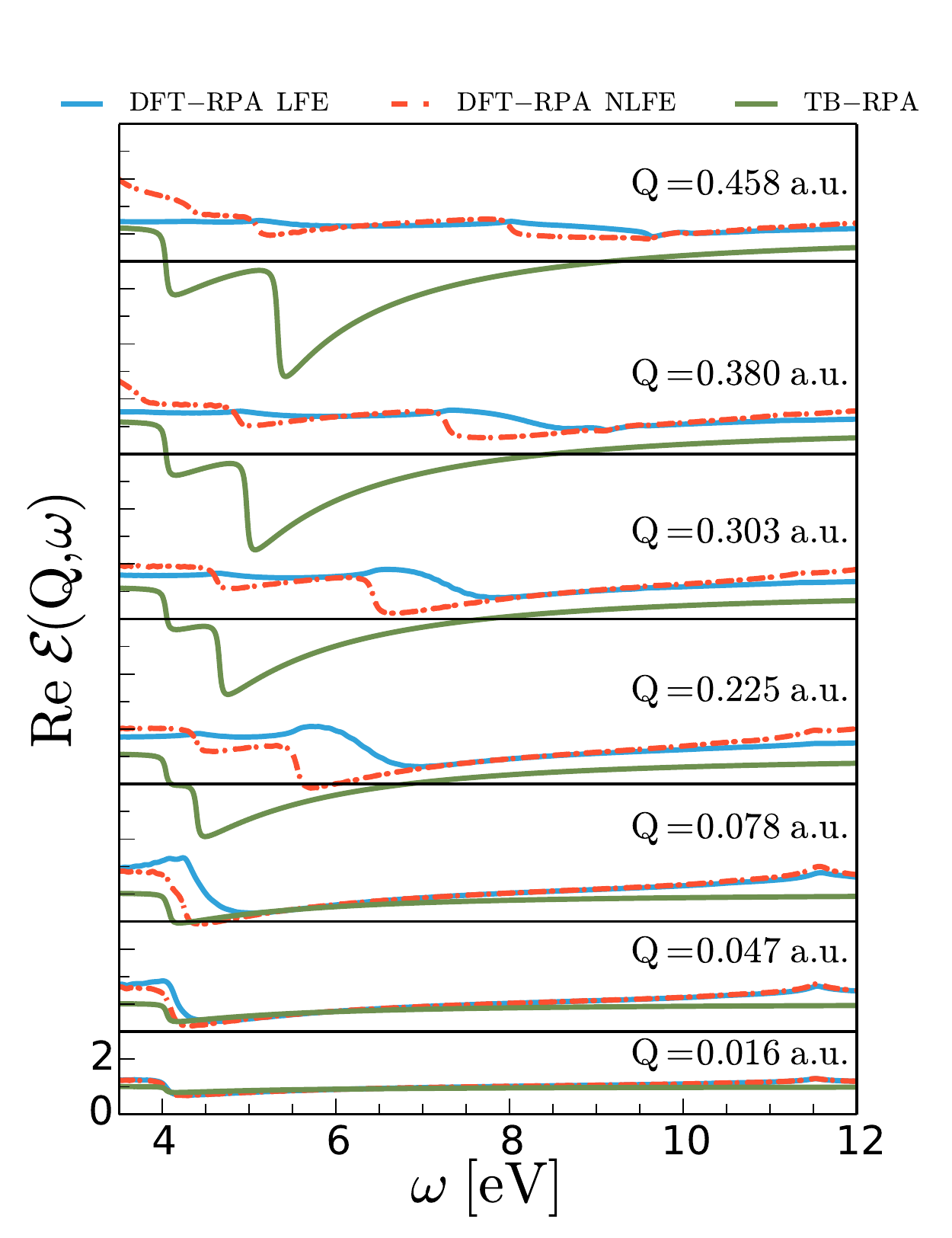}
\caption{Real part of the dielectric function for a several Q wavevectors. Blue line represents the reults obtained by DFT-RPA method with LFE included ($\mathrm{Re\ {\cal E}_M}$), red dashed-dotted line without LFE ($\mathrm{Re\ {\cal E}_{00}}$), and the green line represents the TB-RPA results ($\mathrm{Re\ \varepsilon}$). Zero values of $\mathrm{Re\ {\cal E}}$ are denoted by black horizontal lines.}  
\label{Fig5}
\end{figure}

Fig. \ref{Fig1} shows numerical results for $\mathrm{{\cal E}_1(Q,\omega)}$, $\mathrm{{\cal E}_2(Q,\omega)}$, $\mathrm{D(Q,\omega)}$ and $\mathrm{S(Q,\omega)}$ for six different values of Q chosen in the $\mathrm{\Gamma M}$ direction of the BZ. For small Q values we can notice the absence of screening, i.e. $\mathrm{D(Q,\omega)\approx1}$, for almost all $\omega$ values. 
Pronounced peaks at $4.1\ \mathrm{eV}$ and $\mathrm{13.9\ eV}$ correspond to $\pi\rightarrow\pi^{\ast}$ and $\sigma\rightarrow\sigma_1^{\ast}$ transitions, respectively, around the M point of BZ. As we can see, at these energies the screening factor $\mathrm{D(Q,\omega)}$ is exactly 1. This means that the mentioned transitions are not screened, i.e. the peaks in $\mathrm{S(Q,\omega)}$ are pure SP excitations which appear at the same energies as peaks in $\mathrm{{\cal E}_2(Q,\omega)}$ \cite{Nanolett,Kupcic}. Blue dots in Figs. \ref{Fig3}a and \ref{Fig3}b show the energies of the peaks in $\mathrm{{\cal E}_2(Q,\omega)}$, and red dots the energies of the peaks in $\mathrm{S(Q,\omega)}$ as functions of the wavevector Q. Blue points show characteristic $\mathrm{Q^2}$ dispersion of $\pi$ and $\sigma$ SP transitions. It can be clearly seen that for small Q peaks in $\mathrm{{\cal E}_2(Q,\omega)}$ and peaks in $\mathrm{S(Q,\omega)}$ coincide and follow the same $\mathrm{Q^2}$ dependence which confirms their SP character. This quadratic dispersion of SP excitations is a result of the $\pi$ and $\sigma$ band structure around the saddle point M, as sketched in Fig. \ref{Fig2}. As Q increases the screening factor $\mathrm{D(\mathrm{Q},\omega)}$ increases, enhancing the spectral weight of the peaks and moving them to higher energies, i.e. away from the initial $\pi\rightarrow\pi^{\ast}$ and $\sigma\rightarrow\sigma^{\ast}$ energies (Figs. \ref{Fig1}b,c). This is also visible in Figs. \ref{Fig3}a and \ref{Fig3}b which represent the gradual modification of the initial SP transitions into collective excitations as the dynamical screening becomes more efficient. In this regime ($\mathrm{Q\gtrsim0.03\ a.u.}$) one can with confidence treat these excitations as $\pi$ and $\pi+\sigma$ plasmons, though their broad spectral shapes indicate the presence of Landau damping.
Diamonds in Fig. \ref{Fig3}a show the energies of $\pi$ plasmon peaks in the measured spectra \cite{Pipl1}. We see very nice agreement with our theoretical calculation throughout the whole energy region. Therefore, the pronounced spectral structures which appear for $\mathrm{Q}\approx0$ (e.g. in optical absorption spectra) have purely SP character, while spectral structures which appear for finite Q (e.g. in EELS) represent collective excitations or plasmons.
For $\mathrm{Q\approx0.1\ a.u.}$ the screening becomes most efficient and $\mathrm{{\cal E}_1(Q,\omega)}$ approaches its lowest value.  
Accordingly, in Fig. \ref{Fig3}a we can see the largest shift of the $\pi$ plasmon energy compared to the $\pi\rightarrow\pi^{\ast}$ transition energy. 
For even higher wavevectors Q this shift becomes smaller, the plasmon energy slowly approaches the unscreened $\pi\rightarrow\pi^{\ast}$ transition energies, but the shape of the plasmon peak remains almost unchanged (Figs. \ref{Fig1}e and \ref{Fig1}f).

These effects can also be observed in Fig. \ref{Fig5} which shows energy dependence of the real part of the dielectric function obtained with DFT-RPA and TB-RPA methods. In the TB-RPA approximation where only $\pi$ electrons participate in the screening $\mathrm{\varepsilon_1(Q,\omega)}$ crosses zero for all Q above some minimum value where the resulting dispersion relation reaches the SP continuum, as shown in Fig. \ref{Fig3}c. So, it is obvious that the $\pi$ electrons in TB-RPA for higher Q's behave like a 2D electron gas showing the $\mathrm{\sqrt{Q}}$ dispersion relation. However, for $\mathrm{Q\rightarrow0}$ the excitation energy approaches the finite value at the upper boundary of the $\pi\rightarrow\pi^{\ast}$ continuum ($\approx 4\ \mathrm{eV}$) with the $\mathrm{Q^2}$ dependence (red curve in Fig. \ref{Fig3}c), and does not follow the $\mathrm{\sqrt{Q}}$ line to zero (orange curve in Fig. \ref{Fig3}c). The reason for this is high interband Landau damping and general reduction of macroscopic screening for finite systems in the low Q region \cite{graphite2,macro}. This is all in qualitative agreement with our resluts, but we can conclude that $\mathrm{\sqrt{Q}}$ dispersion is obtained only if we neglect other electronic transitions in graphene. So in the DFT-RPA calculation, including full graphene band structure, i.e. $\sigma$ electrons, a number of zeros of $\mathrm{{\cal E}_1(Q,\omega)}$ is strongly reduced, and if we further include finite lattice effects, i.e. LFE, $\mathrm{{\cal E}_1(Q,\omega)}$ never cross zero in the whole $\mathrm{(Q,\omega)}$ range (Fig. \ref{Fig5}). The spectra still shows well defined though broader peaks, but the plasmon energies are in a much better agreement with the experimental results than the TB-RPA calculation. Also, for very large Q the $\pi$ plasmon energy should approach the SP $\pi\rightarrow\pi^{\ast}$ transition energy, which is not achieved in TB-RPA.

It might seem that our TB-RPA results qualitatively agree with the theoretical interpretation of the measured data in  Ref. \cite{Pipl1}. There the authors demonstrated that the graphene $\pi$ plasmons represent the in-plane charge density oscillations which led them to the conclusion that they behave like plasmons in a 2D electron gas with $\sim\sqrt{\mathrm{Q}}$ dispersion relation. Now we clearly 
see that this conclusion is indeed partially true, but at the same time we see its limitations. First of all, even the TB-RPA results confirm $\sim\sqrt{\mathrm{Q}}$ dispersion only for bigger Q's, while 
the authors missed the fact that for small Q's it shows a $\mathrm{Q^2}$ dependence. This is due to absence of dynamical screening and collective mode changes its character to single particle 
excitations. Possible reason for this failure is because the first measured nonzero Q point is at $\mathrm{Q\approx0.05\ a.u.}$ (or $\mathrm{0.1\ \AA^{-1}}$) above the 
$\mathrm{Q^2}$ region, as can be clearly seen by observing the authors results in Fig. \ref{Fig3}a. So after inclusion of all other points, for $\mathrm{Q\gtrsim0.05\ a.u.}$, one gets the impression that 
the dispersion has a square root behavior. 

Apart from these deviations for small Q's there are other arguments which violate this simple 2D plasma attribution. As we already mentioned, the inclusion of realistic crystal and band structure ($\sigma$ bands and LFE) substantially modifies the TB-RPA  $\sim\sqrt{\mathrm{Q}}$ dispersion which for bigger Q's even becomes linear. This is also confirmed 
by the excellent agreement of our DFT-RPA dispersion and the experimental data shown in Fig. \ref{Fig3}a.

The dispersion of the so-called $\pi+\sigma$ plasmon is more complicated, as visible in Figs. \ref{Fig1} and \ref{Fig3}b. There are in fact not one, but two modes, one originating from $\sigma\rightarrow\sigma_1^{\ast}$ transitions ($\sigma_1$ plasmon) and the other from $\sigma\rightarrow\sigma_2^{\ast}$ transitions ($\sigma_2$ plasmon), as shown in Fig. \ref{Fig2}. For low Q values their dispersion is shown in Fig. \ref{Fig3}b, together with energies of unscreened SP transitions. While they generally follow the behaviour described above for $\pi$ plasmons, from SP excitations to the Landau damped plasmons and back, they give much broader structures, and furthermore there are additional features arising from their mutual interference. For small Q $\sigma\rightarrow\sigma_1^{\ast}$ excitation dominates in intensity, but around $\mathrm{Q\approx0.08\ a.u.}$ it becomes modified by $\sigma\rightarrow\sigma_2^{\ast}$ transitions which gain spectral weight, so the high energy spectra of graphene are dominated by $\sigma_2$ plasmons. For large $\mathrm{Q\gtrsim0.4\ a.u.}$ values the spectra again show unscreened SP excitations.

Therefore, as for the $\pi$ plasmon, in Ref. \cite{Pipl1} the authors make a hasty conclusion about 
the $\sqrt{\mathrm{Q}}$ dispersion of $\sigma$ plasmon, also not taking into account that two kinds of $\sigma\rightarrow\sigma^*$ excitations exist close in energy and influence each other. In experimental papers Ref. \cite{Pipl21,Pipl2} the authors also obtained $\pi$ plasmon dispersion relation which is in accordance with our theoretical results and conclusions, however they are also prone to describe it as $\sqrt{\mathrm{Q}}$ behaviour, with a help of the hydrodinamic model \cite{hidro1,hidro2}.

Our above analysis partially agrees with the results recently published in Ref. \cite{Nanolett}, though not entirely, since its authors claim that the $\pi$ and $\pi+\sigma$ plasmons in graphene do not exist at all. Namely, they derived their conclusions by analysing the dielectric function in the $\mathrm{Q\rightarrow 0}$ (optical) limit which is indeed the region where the $\pi$ and $\sigma$ excitations behave as unscreened SP transitions. However, as we showed, this is not the case for higher Q's. One of the methods to demonstrate that $\pi$ and $\pi+\sigma$ excitations represent self-sustaining charge density oscillations is to induce them by some external perturbation and see how they behave. If these charge oscillations survive at least one period, after the perturbation is being switched off, then they represent collective modes. In Ref. \cite{Osc1} the charge density oscillations in graphene are driven by suddenly created point charge. The authors have identified two periods of oscillations whose frequencies are associated with $\pi$ and $\pi+\sigma$ plasmon frequencies. In Ref. \cite{Osc2} the charge  density in graphene is driven by a point charge moving with constant velocity parallel to the graphene surface. There appear several rows of bow waves whose wavelength, depending on the speed of the point charge, was associated with the excitations of $\pi$ or $\pi+\sigma$ plasmons. These two observations undoubtedly confirm that $\pi$ and $\pi+\sigma$ plasmons indeed exist.

A brief analytic discussion could help to clarify the shape of numerically calculated spectra. The spectrum
\begin{equation}
\mathrm{S(Q,\omega)={\cal E}_2(Q,\omega)D(Q,\omega)}
\label{spec3}
\end{equation}
in the absence of screening ($\mathrm{D=1}$) will have maxima at the energies of $\pi\rightarrow\pi^{\ast}$ and $\sigma\rightarrow\sigma^{\ast}$ SP transitions, as in Fig. \ref{Fig1}a, with $\mathrm{Q^2}$ dispersion coming from the shape of the $\pi$ and $\sigma$ bands. As the screening term $\mathrm{D(Q,\omega)}$ increases with larger Q, these SP peaks will be enhanced and shifted to the maxima $\Omega$ of this term, given by the condition
\begin{equation}
\frac{d}{d\omega}\mathrm{D(Q,\omega)}\Big|_{\omega=\Omega}=0,
\end{equation}
or explicitly
\begin{equation}
\mathrm{{\cal E}_1(Q,\Omega)+\frac{{\cal E}_2(Q,\Omega){\cal E}_2'(Q,\Omega)}{{\cal E}_1'(Q,\Omega)}=0},
\label{cond1}
\end{equation}
where the prime denotes the derivative with respect to $\omega$. This systematic transition from the SP peaks to the maxima of the screening function can be seen in Figs. \ref{Fig1}b-\ref{Fig1}f, and are the proof of the increasingly collective character of these excitations for higher Q. At the same time the dispersion deviates from the initial $\mathrm{Q^2}$ dependence, as seen in Figs. \ref{Fig3}a and \ref{Fig3}b. From (\ref{cond1}) we can explicitly see that the peak positions are not given by the condition 
\begin{equation}
\mathrm{{\cal E}_1(Q,\Omega)}=0,
\label{cond2}
\end{equation}
as would be expected for an ideal collective mode, and indeed, in our full calculation including LFE (\ref{cond2}) is not satisfied for any $\mathrm{(Q,\omega)}$. Nevertheless we find broad but well defined spectral peaks corresponding to plasmons in graphene, though restricted to the region of finite wavevectors and Landau damped. We can also derive the condition (\ref{cond1}) assuming that the spectrum (\ref{spec3}) has a resonant form, with the maximum at the complex pole of the dielectric function:
\begin{equation}
\mathrm{{\cal E}(Q,\Omega-}i\mathrm{\Gamma(Q))}=0,
\label{cond3}
\end{equation}
by expanding around $\Omega$ with $\Gamma\ll\Omega$.

Dielectric function near the resonance energies $\Omega$ shows an interesting behaviour. By inspection of numerical results for finite Q we see
\begin{eqnarray}
\mathrm{{\cal E}_1(Q,\Omega)}&\approx &\mathrm{{\cal E}_2(Q,\Omega)>0},\nonumber\\
\mathrm{{\cal E}_1'(Q,\Omega)}&\approx &\mathrm{-{\cal E}_2'(Q,\Omega)>0},
\label{cond4}
\end{eqnarray}
which is also in agreement with the condition (\ref{cond1}), valid in the resonance or plasmon region. So the properties (\ref{cond4}) can be connected with the collective character of the excitations in this region.

\section{\label{sec:4} Conclusion}
In this paper we have presented the method to determine the character of electronic excitations in solids and applied it to the $\pi$ and $\sigma$ electron excitations in a single layer of pristine graphene. Analysing the energy and momentum dependence of the dielectric function components, which determine the excitation spectra, and especially the dynamical screening factor $\mathrm{D(Q,\omega)}$, we have demonstrated that $\pi\rightarrow\pi^{\ast}$ and $\sigma\rightarrow\sigma^{\ast}$ transitions show two types of behaviour. For small wavevector $\mathrm{Q\approx0}$ they indeed behave like pure interband (SP) transitions, because the screening is completely absent, but as Q increases and the screening becomes prominent, they rather suddenly acquire a collective character. In this Q region one can indeed say that $\pi$ and $\pi+\sigma$ plasmons exist in graphene, though they are always Landau damped and appear as broad structures in the spectra. 
We also demonstrated that because of strong Landau damping, the dielectric function never crosses zero, even in the region of collective 
excitations. These conclusions, based on exact numerical results, are also confirmed by a brief analytic discussion.

We have partially confirmed the seemingly conflicting results of Ref. \cite{Nanolett} and Refs. \cite{Pipl1, Pipl2}. The claim in \cite{Nanolett} that $\pi$ and $\sigma$ excitations are unscreened SP transitions is indeed correct in the low Q region. This result is expected because in 2D materials, like graphene, macroscopic screening 
$\mathrm{{\cal E}_1}$ is significantly reduced in comparison with the bulk system like graphite \cite{graphite2,macro}.
Our results agree very well with the experimental data of Ref. \cite{Pipl1}, however we partially disagree that $\pi$ and $\sigma$ electrons behave as 2D electrons 
with $\mathrm{\sqrt{Q}}$ dispersion relation. We showed that $\pi$ and $\sigma$ excitations change their character: in $\mathrm{Q\approx0}$ region they have SP character with $\mathrm{Q^2}$ dispersion, as Q increases they acquire collective character with $\mathrm{\sqrt{Q}}$ like dispersion, and for even higher Q's they again acquire SP 
character with linear dispersion. In this way we have presented a complete description of this quite intriguing and previously controversial problem, explaining the changing character of electronic excitations in graphene.

By presenting our conclusions we would like to emphasize once again the importance of the in-depth interpretation of experimental measurements and how they relate to the theoretical results. So, to understand the nature of  plasmon dispersion it is crucial to perform measurements for all values of the wavevector Q (e.g. with optical absorption spectra and EELS) and to give a careful theoretical interpretation in each of the limits (i.e. $\mathrm{Q\approx0}$ and $\mathrm{Q\neq0}$). 

\begin{acknowledgements}
The authors are grateful to Donostia International Physics Center (DIPC) and Pedro M. Echenique for hospitality and financial support during various stages of this work. We also thank A. Lucas, 
I. Kup\v ci\'c and V. Silkin for useful discussions. Computational resources were provided by the DIPC computing center.
\end{acknowledgements}


\begin{thebibliography}{99}
\bibitem{gamma0} We have chosen $\gamma\approx2.02\ \mathrm{eV}$ value so that the spectra of TB-RPA around $\mathrm{Q\approx0}$ are comparable with our results obtained with DFT-RPA.
\bibitem{q0} When we have $G_z=G_z'=0$, the point $\mathrm{Q}=0$ needs to be treated with care because then the Coulomb interaction (\ref{3Dexm}) is divergent. This can be resolved by expanding the matrix elements (\ref{Matrel}) for small Q. With this we get additional Q term in front of the interband matrix element which cancels the divergent $\mathrm{Q^{-2}}$ term in the Coulomb interaction when we calculate the dielectric tensor (\ref{epsm}).
\end{thebibliography}
\end{document}